%% file: main.tex
\documentclass[journal]{IEEEtran}

\usepackage{cite}
\usepackage{amsmath,amssymb,amsfonts}
\usepackage{graphicx}
\usepackage{booktabs}
\usepackage{algorithm}
\usepackage{algorithmic}
\usepackage{textcomp}
\usepackage{xcolor}

\usepackage{dblfloatfix}
\usepackage{cuted}
\usepackage{placeins}
\usepackage{pgfplots}
\pgfplotsset{compat=1.18}

\usepackage{caption}

\usepackage{pgfplots}
\pgfplotsset{compat=1.18}
\usepackage{tikz}
\usepackage{pgfplots}
\definecolor{ForestGreen}{RGB}{34,120,70}
\pgfplotsset{compat=1.18}
\DeclareRobustCommand{\circlednum}[1]{%
    \tikz[baseline=(char.base)]{
        \node[
            shape=circle,
            fill=black,
            text=white,
            inner sep=1pt
        ] (char) {\fontsize{7pt}{7pt}\selectfont #1};
    }%
}

\newcommand{\supc}[1]{%
    (\tikz[baseline=(char.base)]{
        \node[shape=circle, fill=black, text=white, inner sep=0.5pt] (char) {\small #1};
    })
}
\begin{document}

\title{\texttt{FloodReasonBench:} Benchmarking VLM Reasoning Segmentation for Embodied Flood Response at the Edge}


\title{\texttt{FloodReasonBench:} Benchmarking VLM Reasoning Segmentation for Embodied Flood Response at the Edge}

\author{
Rajat Bhattacharjya\IEEEauthorrefmark{2}*,
Yoomee Jung\IEEEauthorrefmark{4},
Minwoo Kim\IEEEauthorrefmark{4},
Sing-Yao Wu\IEEEauthorrefmark{2},\\
Eli Bozorgzadeh\IEEEauthorrefmark{2},
Nalini Venkatasubramanian\IEEEauthorrefmark{2},
Nikil Dutt\IEEEauthorrefmark{2}
\\
\IEEEauthorrefmark{2} University of California, Irvine
; \IEEEauthorrefmark{4} Kookmin University, South Korea
\\
*\small Corresponding author: rajatb1@uci.edu
\thanks{Paper is currently under review. Authors' version posted for personal use and not for redistribution. The dataset and code will be made public upon acceptance.}
}

\IEEEaftertitletext{\vspace{-4ex}}
\maketitle
\begin{abstract}
Reasoning segmentation enables vision-language models (VLMs) to translate mission-relevant language requests into pixel-level visual grounding, offering a natural perception interface for embodied agents.
However, existing benchmarks largely focus on generic visual scenes and overlook the domain and resource constraints encountered in flood-response platforms.
We present \texttt{FloodReasonBench}, a benchmark for VLM reasoning segmentation for embodied flood response at the edge. 
At its core, \texttt{FloodReasonBench} introduces \texttt{FloodResponseSeg}, a flood-specific reasoning-segmentation dataset constructed from real-world scenes and response-relevant targets. 
Beyond task accuracy, the benchmark characterizes reasoning-segmentation pipelines under lightweight visual encoding, hierarchical split inference, and compressed intermediate representations. 
We observe strong partition-dependent accuracy variation in the generic pre-adaptation setting, while the flood-adapted target-workload design space exhibits a substantially more compact accuracy range across partitions. 
Evaluation on an NVIDIA Jetson AGX Xavier further exposes the tradeoffs among reasoning-segmentation accuracy, edge-side latency, energy, and communication footprint, enabling quality-constrained selection of edge operating points. 
Together, these results provide a task- and system-level characterization of reasoning segmentation for resource-constrained embodied flood response at the edge.
\end{abstract}


\input{intro}

\input{prob}

\input{method}

\input{expt}

\input{eval}
\input{conclusion}

\begingroup
\providecommand{\BIBdecl}{}
\renewcommand{\BIBdecl}{\scriptsize}
\bibliographystyle{IEEEtran}
\bibliography{ref_full}
 \endgroup

\end{document}

%% file: intro.tex
\section{Introduction}

\begin{figure}[t]
    \centering
    \includegraphics[width=\linewidth]{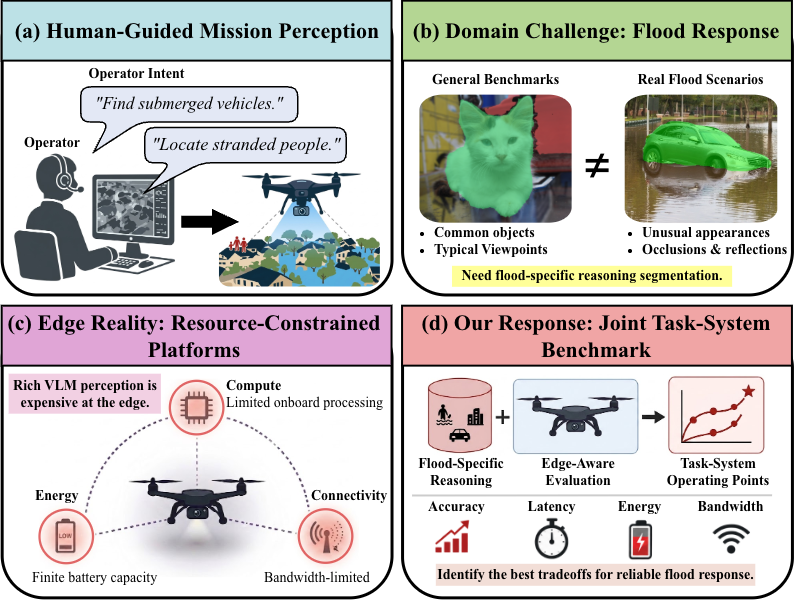}\captionsetup{
    font=footnotesize,
    labelfont={bf,footnotesize}
}
    \caption{Motivation and scope of \texttt{FloodReasonBench}.
    (a) Human responders specify mission-relevant information needs in natural language to embodied platforms operating in flood environments.
    (b) Generic reasoning-segmentation benchmarks do not capture the visual conditions encountered in real flood-response scenarios, motivating a flood-specific reasoning workload.
    (c) Embodied platforms must support such perception under constrained compute, energy, and connectivity.
    (d) \texttt{FloodReasonBench} jointly evaluates task accuracy and system-level resource costs to identify operating points that satisfy deployment-specific quality and resource requirements.}
    \label{fig:motivation}
    \vspace{-3ex}
\end{figure}

Flood-response operations increasingly rely on autonomous platforms such as unmanned aerial vehicles (UAVs) and mobile robots to acquire timely situational awareness in environments that may be unsafe or difficult for human responders to access~\cite{bhattacharjya2025avery}. 
For such embodied agents, perception must go beyond recognizing objects in an image: the system must associate a mission-relevant request from a human responder or operator with the corresponding regions in the physical environment.
Vision-language models (VLMs) coupled with reasoning segmentation provide a natural interface for this interaction by grounding language-specified intent into pixel-level regions~\cite{bhattacharjya2025avery}. 
For example, an operator may ask an aerial platform to identify people requiring attention, locate partially submerged vehicles, or segment affected buildings from a flood scene. 
Such perception, however, must operate under the compute, energy, and communication constraints of the embedded platform~\cite{bhattacharjya2025avery,memoguard} as summarized in Fig.~\ref{fig:motivation}. 

Existing reasoning-segmentation benchmarks~\cite{lisa} largely focus on generic visual scenes and do not capture the domain-specific conditions encountered during flood response. 
This domain gap is consequential: without flood-specific fine-tuning, the VLM LISA~\cite{lisa} achieves 0.7223 gIoU and 0.7385 cIoU on our flood-response workload; whereas flood-specific adaptation improves these to 0.8423 and 0.9013, corresponding to gains of 12.0 and 16.28 points, respectively. 
Deployment introduces a second challenge. LISA relies on the heavyweight SAM~\cite{kirillov2023segment} image encoder, making fully onboard execution costly in compute and energy~\cite{bhattacharjya2025avery}, while fully remote execution increases communication requirements and dependence on connectivity. 
Split computing~\cite{splitz} offers a middle ground by executing part of the model onboard and transmitting an intermediate representation for remote continuation. 
Yet, in hierarchical visual backbones, candidate partition points expose representations with different dimensions, semantic content, and sensitivity to compression. 
Choosing where to split therefore depends not only on how much computation is placed at the edge, but also on whether the exposed representation preserves sufficient task-relevant information and satisfies the required perception quality.

We introduce \texttt{FloodReasonBench}, a task- and system-level benchmark for VLM reasoning segmentation for embodied flood response at the edge. 
The benchmark couples a new flood-specific reasoning-segmentation workload with systematic evaluation of lightweight and partitioned visual inference under embedded-resource constraints. 
In doing so, \texttt{FloodReasonBench} connects two questions that are typically studied separately: \textit{(1) how effectively does reasoning segmentation transfer to the flood-response domain?} and 
\textit{(2) how to partition and deploy this capability under constrained edge and remote resources?}
The benchmark exposes a design space for selecting operating points subject to user-defined perception-quality requirements and available compute, energy, and communication resources.

The main contributions of this work are:
\begin{itemize}
\vspace{-0.7mm}
\item We introduce \texttt{FloodReasonBench}, together with \texttt{FloodResponseSeg}, a new flood-specific reasoning-segmentation dataset constructed through a semi-automated curation, grounding, segmentation, and quality-control pipeline. 
Evaluation with an off-the-shelf reasoning-segmentation model further quantifies the domain gap that motivates flood-specific adaptation.

\item We systematically characterize hierarchical split reasoning
segmentation using MobileSAM~\cite{zhang2023faster} and learned
intermediate-feature compression, revealing strong partition
sensitivity on the generic ReasonSeg~\cite{lisa} workload and a
substantially more compact partition-quality range in the
flood-adapted target-workload design space.

\item We evaluate the resulting split configurations on an NVIDIA Jetson AGX Xavier across multiple platform power modes, capturing different levels of compute and power availability that may arise when perception shares an embodied platform with other onboard functions. 
By jointly considering task quality, edge-side latency, energy, and transmitted feature size, we expose a quality-constrained design space for selecting practical operating points under deployment-specific resource requirements.

\end{itemize}

%% file: prob.tex
\vspace{-2mm}

\section{\texttt{FloodResponseSeg}: Flood-Response Reasoning Segmentation Dataset}

\vspace{-1mm}
We now introduce \texttt{FloodResponseSeg}, the flood-specific reasoning-segmentation dataset underlying \texttt{FloodReasonBench}.
It addresses the gap between existing reasoning-segmentation datasets such as ReasonSeg~\cite{lisa}, which primarily contain generic visual scenes, and flood-vision datasets~\cite{floodnet}, which largely target classification, detection, or conventional segmentation rather than language-conditioned visual grounding. 
The current release considers three categories---people, buildings, and vehicles---corresponding to human assistance, affected infrastructure, and mobility during flood-response operations.
Each sample consists of a flood image, a natural-language query representing a responder's information request, and the pixel-level mask of the corresponding target.

\subsection{Dataset Curation}
\label{sec:dataset_curation}
\label{sec:dataset}

\begin{figure}[t]
    \centering
    \includegraphics[width=\linewidth]{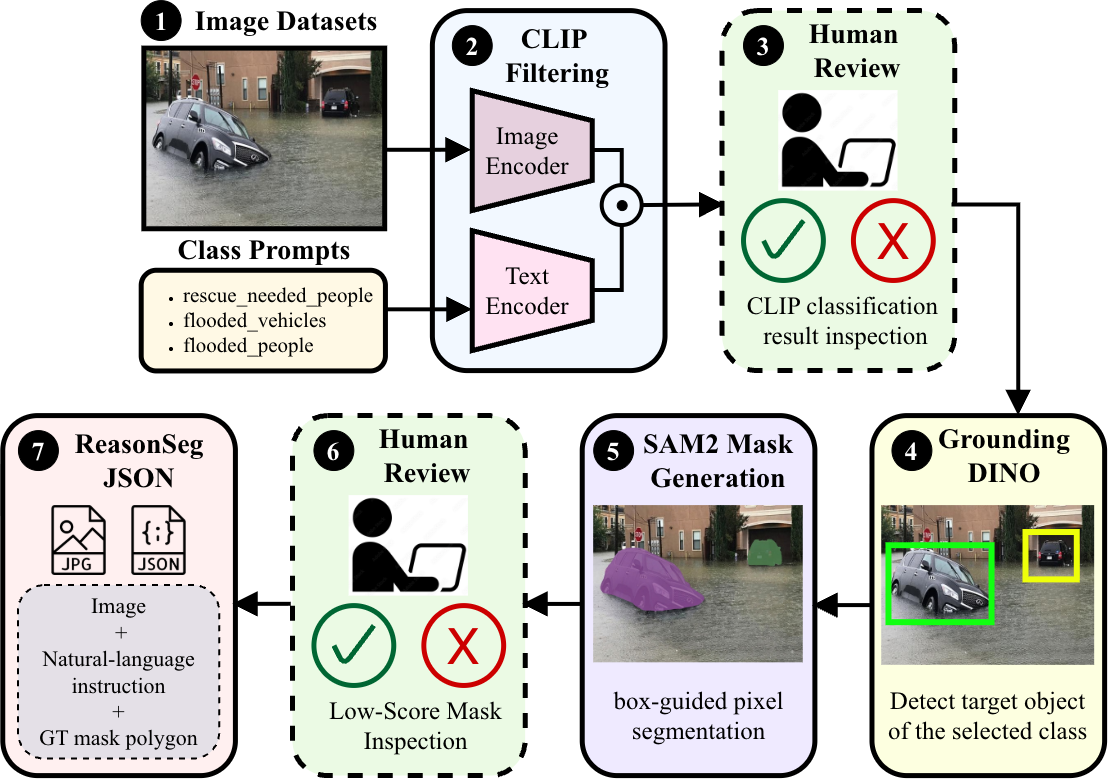}
    \captionsetup{
    font=footnotesize,
    labelfont={bf,footnotesize}
}
    \caption{Semi-automated \texttt{FloodResponseSeg} curation pipeline combining foundation-model-assisted filtering, localization, and mask generation with manual inspection and natural-language query authoring.}
    \label{fig:dataset_pipeline}
\end{figure}

Fig.~\ref{fig:dataset_pipeline} summarizes the semi-automated curation pipeline. 
We first collect real-world flood imagery~\supc{1} and apply CLIP-based~\cite{clip} semantic filtering~\supc{2} to identify images containing the target categories. 
Candidate images are manually inspected~\supc{3}, followed by target localization using Grounding DINO~\cite{liu2024grounding}~\supc{4} and mask generation using SAM2~\cite{sam2}~\supc{5}. 
The generated image--mask pairs undergo manual visual quality inspection, with unsuitable annotations rejected~\supc{6}. 
For each accepted pair, we manually author a natural-language query describing the target in the context of the flood scene~\supc{7}; multiple queries may refer to the same target while sharing the same pixel-level ground truth. 
The resulting image--query--mask triplets are then converted to the reasoning-segmentation format used by LISA~\cite{lisa}.

\vspace{-3mm}

\subsection{Dataset Composition and Augmentation}
\label{sec:dataset_composition}

The training set contains 532 original annotated samples and three photometrically augmented variants per sample, yielding 2,128 training instances. 
Augmentation includes color, brightness, contrast, blur, noise, and sharpening while preserving target geometry and annotations. 
The evaluation set contains 100 non-augmented samples. Table I summarizes the class-wise composition. 

\begin{table}[t]
    \centering
    \captionsetup{
    font=footnotesize,
    labelfont={bf,footnotesize}
}\caption{Composition of the current \texttt{FloodResponseSeg} dataset.}
    \label{tab:dataset_stats}
    \begin{tabular}{lrrrr}
        \toprule
        \textbf{Split} &
        \textbf{People} &
        \textbf{Buildings} &
        \textbf{Vehicles} &
        \textbf{Total} \\
        \midrule
        Train (original)   & 189 & 209 & 134 & 532 \\
        Augmented variants & 567 & 627 & 402 & 1,596 \\
        Evaluation      & 48  & 24  & 28  & 100 \\
        \bottomrule
    \end{tabular}
\end{table}

%% file: method.tex
\vspace{-3mm}
\section{Edge Reasoning-Segmentation Benchmark Design}
\label{sec:edge_design}

\begin{figure*}[t]
    \centering
    \includegraphics[width=\textwidth]{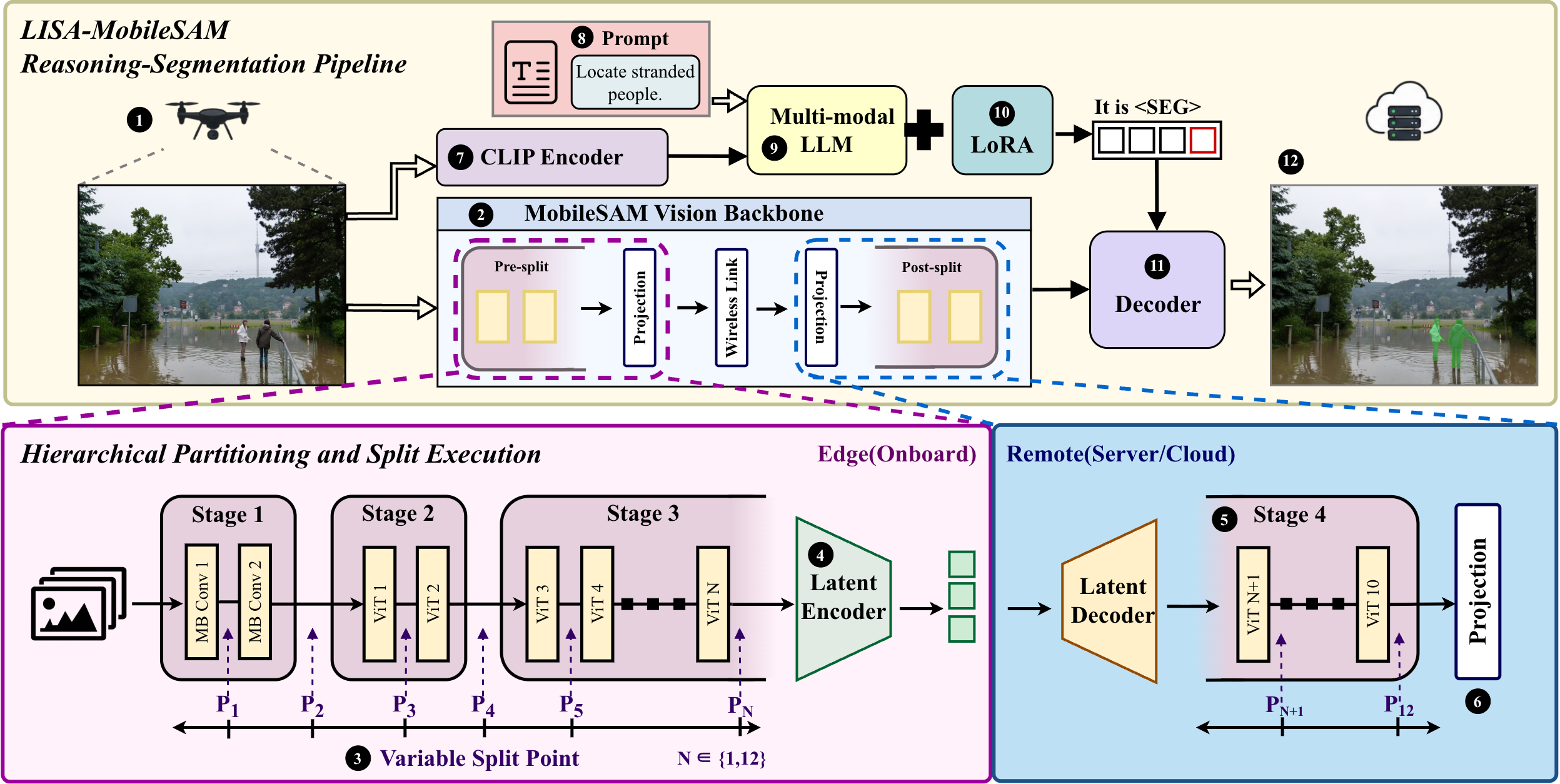}\captionsetup{
    font=footnotesize,
    labelfont={bf,footnotesize}
}
   \caption{Overview of the \texttt{FloodReasonBench} reasoning-segmentation pipeline and hierarchical split-execution design. An embodied platform acquires a flood image~\circlednum{1}, which enters the MobileSAM vision backbone~\circlednum{2}, whose TinyViT partition can be selected across candidate blocks~\circlednum{3}. At the selected cut, the intermediate feature is compressed by a learned latent encoder~\circlednum{4}, transmitted to the remote side, reconstructed, and processed by the remaining backbone~\circlednum{5} and projection~\circlednum{6}. In parallel, the image is processed by the CLIP encoder~\circlednum{7}, while the responder query~\circlednum{8} is interpreted by the multimodal LLM~\circlednum{9} with LoRA adaptation~\circlednum{10}. The resulting \texttt{<SEG>} representation conditions the mask decoder~\circlednum{11} to produce the requested pixel-level segmentation~\circlednum{12}.}
    \label{fig:system_overview}
\end{figure*}

While \texttt{FloodResponseSeg} provides the task-level workload, the systems dimension of \texttt{FloodReasonBench} studies how the reasoning-segmentation pipeline can be structured for resource-constrained edge execution. 
Fig.~\ref{fig:system_overview} summarizes the complete benchmark
design. 
An embodied platform (e.g., a UAV) acquires a flood image~\supc{1},
which enters the MobileSAM visual pathway~\supc{2}--\supc{6},
while its CLIP~\cite{clip} representation~\supc{7} and responder
query~\supc{8} are processed by the multimodal reasoning
pathway~\supc{9}--\supc{10}; the resulting
\texttt{<SEG>} representation drives the mask decoder~\supc{11}
to generate the requested target region~\supc{12}. 
Starting from the original SAM-based LISA~\cite{lisa} configuration as a reference, we consider two increasingly edge-oriented variants: replacing the heavyweight SAM image encoder with MobileSAM, and then partitioning its TinyViT~\cite{tinyvit} backbone with learned intermediate-feature compression. 
These configurations provide the candidate model operating points that are subsequently evaluated jointly with task quality and embedded-system cost.

\vspace{-4.9mm}
\subsection{Lightweight LISA with MobileSAM}
\label{sec:mobile_lisa}

LISA couples multimodal language reasoning with SAM~\cite{kirillov2023segment} to generate the pixel-level mask associated with a natural-language query. 
However, its SAM ViT-H image encoder is computationally expensive for embedded execution~\cite{bhattacharjya2025avery}. 
We therefore replace the SAM image encoder with MobileSAM~\cite{zhang2023faster}, which uses the substantially smaller TinyViT backbone~\cite{tinyvit} while retaining compatibility with the SAM segmentation pathway.

TinyViT has a hierarchical architecture rather than the homogeneous transformer structure of SAM ViT-H. 
In the MobileSAM configuration used here, the backbone contains 12 blocks across four stages: two initial MBConv~\cite{mbconv} blocks followed by ten transformer blocks. 
Throughout the paper, we denote the two MBConv blocks as \emph{MB1} and \emph{MB2}, and index the subsequent transformer blocks as \emph{Blocks 0--9}. 
This hierarchy changes both the amount of computation performed at different depths and the dimensions and characteristics of the intermediate representations exposed for split execution.

We retain the pretrained TinyViT backbone and adapt its feature interface to the LISA segmentation pathway before flood-specific fine-tuning.
The resulting MobileSAM-LISA configuration serves as the lightweight, non-split baseline in \texttt{FloodReasonBench}.
It also exposes a richer split design space than the original SAM-based configuration.
Whereas AVERY~\cite{bhattacharjya2025avery} uses a fixed early partition of the SAM ViT encoder, TinyViT provides structurally distinct candidate cuts throughout its hierarchical backbone.
We therefore study not only the accuracy impact of replacing SAM with MobileSAM, but also how the selected TinyViT partition affects task quality and edge-system cost.

\vspace{-4.0mm}
\subsection{Hierarchical Split Inference and Feature Compression}
\label{sec:hierarchical_split}

The lower portion of Fig.~\ref{fig:system_overview} expands the split execution of the MobileSAM pathway~\supc{2}--\supc{6}. 
For a selected partition~\supc{3}, the embedded platform executes the TinyViT prefix up to that point.
The exposed intermediate feature is compressed by a lightweight latent encoder~\supc{4} and transmitted across the edge--remote boundary. 
A corresponding latent decoder reconstructs the feature remotely, after which execution continues through the remaining TinyViT backbone~\supc{5} and projection~\supc{6}.
This design avoids transmitting the raw intermediate tensor while allowing the amount of visual computation performed onboard to vary with the partition.

We evaluate candidate cuts after each of the 12 TinyViT blocks: MB1, MB2, and transformer Blocks 0--9. 
Because TinyViT is hierarchical, the representations exposed at these locations differ in spatial resolution, channel dimension, and semantic depth. 
Consequently, moving the partition deeper into the backbone does not necessarily reduce the transmitted representation size, and representations from different stages may respond differently to the same compression ratio.
We therefore train a separate autoencoder (AE) for each candidate partition to match the dimensions and characteristics of its corresponding intermediate feature.

The AE serves as an intermediate-feature bottleneck while the TinyViT backbone remains fixed. 
Following the feature-compression setup used in AVERY~\cite{bhattacharjya2025avery}, each AE is trained independently using feature reconstruction before being inserted into the reasoning-segmentation pipeline. 
Keeping the compression module separate from subsequent task adaptation allows us to characterize partition sensitivity under feature compression and subsequently construct split-aware flood-adapted operating points using the same learned compression bottlenecks.

\vspace{-5mm}
\subsection{Split-Aware Flood Adaptation}
\label{sec:split_adaptation}

Intermediate-feature compression introduces reconstruction error before the remaining reasoning-segmentation pipeline. 
The effect of this distortion can vary with the selected TinyViT partition because different stages expose representations with different task-relevant information and feature characteristics. 
We therefore evaluate each MobileSAM partition under two settings: \emph{pre-adaptation} and \emph{split-aware flood adaptation}.



In the pre-adaptation setting, the trained AE is inserted at each candidate partition without flood-specific split adaptation.
We evaluate these configurations on the standard ReasonSeg validation set to characterize the generic partition sensitivity of the hierarchical TinyViT backbone under intermediate-feature compression.

For split-aware flood adaptation, the AE associated with each partition remains fixed while the trainable reasoning-segmentation components are fine-tuned on \texttt{FloodResponseSeg} with the compressed and reconstructed feature in the inference path.
The resulting configurations are then evaluated on \texttt{FloodResponseSeg} to characterize the partition design space for the target flood-response workload. 
Together, these evaluations contrast the generic pre-adaptation partition behavior with the final flood-adapted operating landscape used for edge design-space exploration.

%% file: expt.tex
\vspace{-5mm}
\section{Experimental Methodology}
\label{sec:methodology}

We now evaluate \texttt{FloodReasonBench} along both task- and system-level dimensions. 
At the task level, we characterize flood-domain transfer, the accuracy impact of lightweight visual encoding, and the sensitivity of reasoning segmentation to hierarchical split execution.
At the system level, we characterize the onboard computation and communication footprint associated with each MobileSAM partition on an embedded NVIDIA Jetson AGX Xavier and combine these measurements with task quality to expose deployment operating points.

\vspace{-5mm}
\subsection{Model and Training Setup}
\label{sec:training_setup}

We use LISA-7B~\cite{lisa} as the base reasoning-segmentation model.
The original LISA configuration uses the SAM ViT-H image encoder, while the lightweight configuration replaces it with the MobileSAM TinyViT encoder. 
To first quantify the flood-domain gap, we evaluate the original LISA-SAM checkpoint directly on the \texttt{FloodResponseSeg} evaluation set without flood-specific fine-tuning.
For flood-specific adaptation, the LISA-SAM and LISA-MobileSAM configurations are then fine-tuned on the 2,128-instance \texttt{FloodResponseSeg} training set using a common setup. 
We use a learning rate of 1e-4 and an effective batch size of four. 
The TinyViT backbone remains frozen during MobileSAM-LISA fine-tuning, while the trainable language and segmentation components are adapted to the flood-response workload.

For split execution, a separate AE is trained for each candidate TinyViT partition. 
Following the feature-compression setup of AVERY~\cite{bhattacharjya2025avery}, we train the AEs on ADE20K~\cite{ade20k} by minimizing intermediate-feature reconstruction error while keeping the TinyViT backbone and feature-alignment components fixed. 
Thus, AE optimization is driven by feature reconstruction rather than the downstream segmentation objective. 
ADE20K validation data are used to monitor reconstruction error, while ReasonSeg validation data provide a task-level assessment of the reconstructed representations. 
Unless otherwise stated, the MobileSAM split experiments use a compression ratio (CR) of $0.1$.
We additionally evaluate the original SAM split configuration at CRs of $0.25$, $0.10$, and $0.05$ as a reference.


For split-aware flood adaptation, the AE associated with each partition is loaded and kept fixed while the remaining trainable reasoning-segmentation components are fine-tuned on \texttt{FloodResponseSeg} with the compressed and reconstructed feature in the inference path. 
This preserves the learned compression bottleneck while adapting the downstream task components to the reconstructed split representation encountered during flood-response inference.

\vspace{-3.0mm}
\subsection{Accuracy Evaluation}
\label{sec:accuracy_eval}

We report reasoning-segmentation performance using gIoU and cIoU~\cite{lisa}, following the evaluation protocol of LISA.
gIoU averages IoU over the evaluation samples, whereas cIoU computes the cumulative intersection over the cumulative union across the evaluation set and is consequently more influenced by large-area targets. 
Following LISA, we use gIoU as the primary task-quality metric while reporting cIoU as a complementary measure. 
The non-augmented 100-sample \texttt{FloodResponseSeg} evaluation set is used for flood-domain evaluation. 
We first evaluate the original LISA-SAM checkpoint on this workload without flood-specific fine-tuning to quantify the need for flood-specific adaptation. 
To characterize the generic pre-adaptation partition landscape, we evaluate the split configurations on the standard 200-sample ReasonSeg validation set. 
Following split-aware flood adaptation, the resulting partition configurations are evaluated on \texttt{FloodResponseSeg} to characterize the partition design space for the target flood-response workload.


Our accuracy evaluation proceeds from domain transfer to edge-oriented model design. 
We first compare the original LISA-SAM checkpoint before and after flood-specific adaptation on \texttt{FloodResponseSeg}. 
We then compare the adapted, unsplit LISA-SAM and LISA-MobileSAM configurations. 
Next, we characterize generic MobileSAM partition sensitivity on ReasonSeg and the final split-aware flood-adapted partition landscape on \texttt{FloodResponseSeg}. 
Finally, we jointly analyze the flood-adapted gIoU and cIoU results with embedded latency and energy measurements and compressed representation size. 
For quality-constrained design-space exploration, we use gIoU as the primary quality constraint: given a user-defined minimum gIoU requirement, the benchmark exposes the candidate operating points that satisfy the required perception quality and their corresponding resource costs.

\vspace{-5mm}
\subsection{Embedded Platform and System Metrics}
\label{sec:embedded_eval}

We profile the edge-side execution of the MobileSAM split configurations on an NVIDIA Jetson AGX Xavier.
Inputs are processed at the $1024\times1024$ resolution used by the segmentation pipeline, and the embedded experiments use FP16 inference with CR $=0.1$. 
We sweep multiple supported Jetson power and CPU configurations, including MAXN, 10\,W, 15\,W, and several 30\,W operating modes, to characterize split execution under different onboard resource envelopes. 
This is relevant for embodied platforms such as UAVs, where perception shares finite compute and power resources with other onboard functions such as localization, planning, control, and communication~\cite{access_av, hyperdoa}, and therefore may not always have access to the platform's maximum resources.

For each candidate partition, we measure edge-side latency and average power, compute energy per frame from these measurements, and record the size of the compressed representation transmitted to the remote processor.


Together, edge-side latency and energy quantify the onboard cost of executing the TinyViT prefix and compressing its intermediate representation, while compressed feature size captures the communication footprint associated with each split point. 
These complementary metrics enable \texttt{FloodReasonBench} to characterize the task--compute--communication tradeoffs across the hierarchical partition design space.

%% file: eval.tex
\vspace{-4.0mm}

\section{Results}
\label{sec:results}
\begin{table}[t]
    \centering
\captionsetup{
    font=footnotesize,
    labelfont={bf,footnotesize}
}    \caption{Reasoning-segmentation accuracy on \texttt{FloodResponseSeg}.}
    \label{tab:model_accuracy}
    \begin{tabular}{lccc}
        \toprule
        \textbf{Configuration} & \textbf{CR} & \textbf{gIoU} & \textbf{cIoU} \\
        \midrule
        LISA-SAM (no flood FT)       & --   & 0.7223 & 0.7385 \\
        LISA-SAM (flood FT)          & --   & 0.8423 & 0.9013 \\
        LISA-MobileSAM (flood FT)    & --   & 0.8202 & 0.8572 \\
        \midrule
        SAM Split (flood FT, AVERY~\cite{bhattacharjya2025avery}) & 0.25 & 0.8240 & 0.8533 \\
        SAM Split (flood FT, AVERY~\cite{bhattacharjya2025avery}) & 0.10 & 0.8081 & 0.8304 \\
        SAM Split (flood FT, AVERY~\cite{bhattacharjya2025avery}) & 0.05 & 0.8150 & 0.8547 \\
        \bottomrule
    \end{tabular}
\end{table}

We evaluate \texttt{FloodReasonBench} from both task and system perspectives. 
We first quantify the need for flood-specific adaptation and the accuracy tradeoff introduced by lightweight visual encoding. 
We then characterize the generic pre-adaptation partition landscape on ReasonSeg~\cite{lisa} and the flood-adapted partition landscape on the target \texttt{FloodResponseSeg} workload.
Finally, we combine the flood-adapted task quality with edge-side latency, energy, and compressed representation size to expose quality-constrained deployment tradeoffs.

\vspace{-3mm}
\subsection{Flood-Domain Adaptation and Lightweight Reasoning Segmentation}
\label{sec:results_lightweight}

Table~\ref{tab:model_accuracy} summarizes the primary LISA configurations on \texttt{FloodResponseSeg}.
Without flood-specific fine-tuning, the original LISA-SAM checkpoint achieves 0.7223 gIoU and 0.7385 cIoU. 
Flood-specific adaptation increases these to 0.8423 gIoU and 0.9013 cIoU, corresponding to gains of 12.0 and 16.28 points, respectively. 
Thus, adaptation to the flood-response workload provides substantial gains under both the per-image gIoU and cumulative cIoU metrics.

Replacing SAM ViT-H with MobileSAM after flood-specific adaptation yields 0.8202 gIoU and 0.8572 cIoU, corresponding to reductions of 2.21 gIoU points and 4.41 cIoU points relative to the adapted LISA-SAM configuration. 
The substantially lighter TinyViT visual backbone therefore preserves much of the flood-response reasoning-segmentation accuracy while enabling the hierarchical split design space studied next.

For reference, we additionally evaluate the original SAM split configuration used in AVERY~\cite{bhattacharjya2025avery}.
Across CRs of 0.25, 0.10, and 0.05, the split configuration achieves 0.8081--0.8240 gIoU and 0.8304--0.8547 cIoU, providing a SAM-based compressed split-inference reference for the subsequent MobileSAM study.

\vspace{-3mm}
\subsection{Partition Sensitivity and Split-Aware Flood Adaptation}
\label{sec:results_split}

\begin{figure}[t]
\centering
\begin{tikzpicture}
\begin{axis}[
    width=\linewidth,
    height=0.6\linewidth,
    xlabel={Split Point},
    ylabel={IoU},
    symbolic x coords={MB1,MB2,B0,B1,B2,B3,B4,B5,B6,B7,B8,B9},
    xtick=data,
    xticklabels={MB1,MB2,0,1,2,3,4,5,6,7,8,9},
    x tick label style={font=\scriptsize},
    y tick label style={font=\small},
    ymin=0.50,
    ymax=1.00,
    ytick={0.50,0.55,0.60,0.65,0.70,0.75,0.80,0.85,0.90,0.95,1.00},
    grid=major,
    mark size=1.5pt,
    line width=1pt,
    legend style={
        at={(0.5,0.96)},
        anchor=north,
        legend columns=2,
        font=\scriptsize,
        fill=white,
        draw=black
    },
]

\addplot[
    solid,
    red,
    mark=o,
    mark size=1.5pt
] coordinates {
    (MB1,0.6512)
    (MB2,0.6919)
    (B0,0.6250)
    (B1,0.6315)
    (B2,0.5916)
    (B3,0.5409)
    (B4,0.5437)
    (B5,0.5322)
    (B6,0.6119)
    (B7,0.7146)
    (B8,0.6770)
    (B9,0.6974)
};
\addlegendentry{Pre-adapt. (ReasonSeg) gIoU}

\addplot[
    dashed,
    red,
    mark=o,
    mark size=1.5pt
] coordinates {
    (MB1,0.7522)
    (MB2,0.8149)
    (B0,0.7652)
    (B1,0.7463)
    (B2,0.6854)
    (B3,0.6661)
    (B4,0.6448)
    (B5,0.6473)
    (B6,0.7375)
    (B7,0.8303)
    (B8,0.7759)
    (B9,0.8230)
};
\addlegendentry{Pre-adapt. (ReasonSeg) cIoU}

\addplot[
    solid,
    blue,
    mark=o,
    mark size=1.5pt
] coordinates {
    (MB1,0.7599)
    (MB2,0.7832)
    (B0,0.7695)
    (B1,0.7941)
    (B2,0.7701)
    (B3,0.7774)
    (B4,0.7688)
    (B5,0.7697)
    (B6,0.7696)
    (B7,0.7985)
    (B8,0.7925)
    (B9,0.8025)
};
\addlegendentry{Flood-adapt. gIoU}

\addplot[
    dashed,
    blue,
    mark=o,
    mark size=1.5pt
] coordinates {
    (MB1,0.8309)
    (MB2,0.7840)
    (B0,0.8265)
    (B1,0.8128)
    (B2,0.8259)
    (B3,0.8341)
    (B4,0.8275)
    (B5,0.8294)
    (B6,0.8329)
    (B7,0.8094)
    (B8,0.8236)
    (B9,0.8386)
};
\addlegendentry{Flood-adapt. cIoU}

\end{axis}
\end{tikzpicture}
\captionsetup{
    font=footnotesize,
    labelfont={bf,footnotesize}
}
\vspace{-6mm}
\caption{Reasoning-segmentation accuracy across MobileSAM split points at CR $=0.1$. Pre-adaptation configurations are evaluated on ReasonSeg to characterize generic partition sensitivity, while split-aware flood-adapted configurations are evaluated on \texttt{FloodResponseSeg}. The pronounced partition-dependent variation observed in the generic setting is substantially less evident in the flood-adapted target-workload design space.}
\label{fig:adaptation_iou}
\vspace{-4mm}
\end{figure}
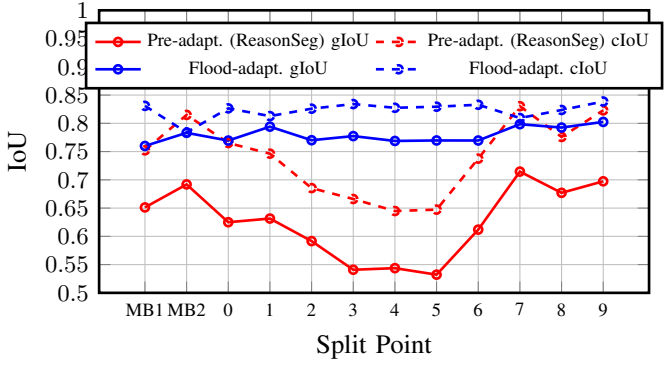

We next examine reasoning-segmentation quality across all 12 MobileSAM partition points at CR $=0.1$. 
Fig.~\ref{fig:adaptation_iou} first shows the pre-adaptation split configurations evaluated on ReasonSeg, providing a reference for the generic partition sensitivity of the hierarchical TinyViT backbone. 
gIoU varies from 0.5322 to 0.7146 across the partitions, an 18.24-point spread, while cIoU varies from 0.6448 to 0.8303, an 18.55-point spread. 
The pronounced non-monotonic variation shows that split-compression performance is strongly dependent on which hierarchical TinyViT representation is exposed at the partition.



Following split-aware flood adaptation, we characterize all 12 partition configurations on \texttt{FloodResponseSeg}, representing the target flood-response workload.
Despite spanning structurally different locations in the TinyViT hierarchy, the resulting gIoU values remain within 0.7599--0.8025, a 4.26-point spread, while cIoU ranges from 0.7840 to 0.8386, a 5.46-point spread. 
In contrast to the pronounced partition-dependent variation observed in the generic ReasonSeg reference, the final flood-adapted design space exhibits a comparatively compact accuracy range across split points.

Consequently, task quality alone does not strongly distinguish many of the candidate partitions in the target deployment setting. 
Multiple TinyViT cuts achieve comparable reasoning-segmentation accuracy, leaving edge latency, energy, and compressed representation size as important dimensions for selecting among quality-feasible operating points.
We examine these system-level tradeoffs next.

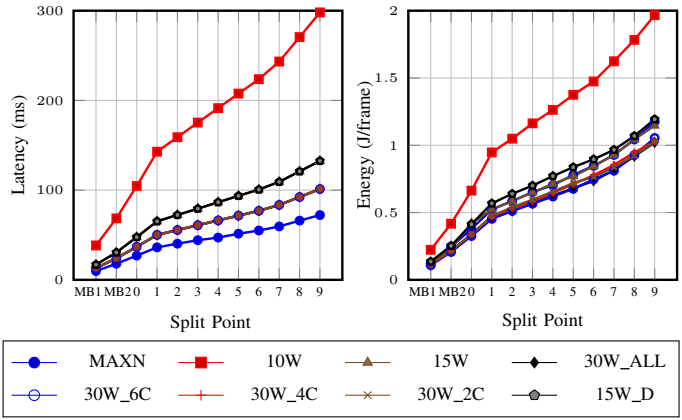
\begin{figure}[t]
\centering

\begin{tikzpicture}

\begin{axis}[
    name=latency,
    at={(0,0)},
    anchor=south west,
    width=0.58\linewidth,
    height=0.58\linewidth,
    xlabel={Split Point},
    ylabel={Latency (ms)},
    xlabel style={font=\scriptsize, xshift=1pt},
    ylabel style={font=\scriptsize, yshift=-4pt},
    symbolic x coords={10,11,0,1,2,3,4,5,6,7,8,9},
    xtick=data,
    xticklabels={\hspace{-6pt}MB1,\hspace{1pt}MB2,0,1,2,3,4,5,6,7,8,9},
    x tick label style={font=\tiny, yshift=1pt},
    y tick label style={font=\tiny},
    ymin=0,
    ymax=300,
    grid=major,
    mark size=1.4pt,
    line width=0.9pt,
]

\addplot+[mark=*, solid] coordinates {
(10,9.8) (11,17.9) (0,27.0) (1,36.2)
(2,40.3) (3,44.2) (4,47.1) (5,51.4)
(6,55.0) (7,59.5) (8,66.0) (9,72.2)
};

\addplot+[mark=square*, solid] coordinates {
(10,38.3) (11,68.5) (0,104.6) (1,142.8)
(2,159.0) (3,175.3) (4,191.4) (5,207.6)
(6,223.7) (7,243.3) (8,270.6) (9,298.1)
};

\addplot+[mark=triangle*, solid] coordinates {
(10,16.9) (11,30.6) (0,47.7) (1,65.3)
(2,72.4) (3,79.4) (4,86.5) (5,93.6)
(6,100.7) (7,109.3) (8,121.0) (9,132.5)
};

\addplot+[mark=diamond*, solid] coordinates {
(10,13.2) (11,24.0) (0,36.9) (1,50.1)
(2,55.5) (3,60.9) (4,66.3) (5,71.5)
(6,77.0) (7,83.5) (8,92.3) (9,101.3)
};

\addplot+[mark=o, solid] coordinates {
(10,13.2) (11,23.9) (0,36.8) (1,50.1)
(2,55.5) (3,60.9) (4,66.2) (5,71.6)
(6,76.9) (7,83.5) (8,92.3) (9,101.2)
};

\addplot+[mark=+, solid] coordinates {
(10,13.2) (11,24.0) (0,36.9) (1,50.1)
(2,55.5) (3,60.9) (4,66.2) (5,71.6)
(6,77.0) (7,83.5) (8,92.4) (9,101.2)
};

\addplot+[mark=x, solid] coordinates {
(10,13.2) (11,23.9) (0,36.9) (1,50.1)
(2,55.5) (3,60.9) (4,66.3) (5,71.6)
(6,76.9) (7,83.5) (8,92.4) (9,101.3)
};

\addplot+[mark=pentagon*, solid] coordinates {
(10,16.9) (11,30.5) (0,47.6) (1,65.2)
(2,72.3) (3,79.3) (4,86.5) (5,93.6)
(6,100.6) (7,109.2) (8,120.9) (9,132.6)
};

\end{axis}

\begin{axis}[
    name=energy,
    at={(0.50\linewidth,0)},
    anchor=south west,
    width=0.58\linewidth,
    height=0.58\linewidth,
    xlabel={Split Point},
    ylabel={Energy (J/frame)},
    xlabel style={font=\scriptsize, yshift=1pt},
    ylabel style={font=\scriptsize, yshift=-7pt},
    symbolic x coords={10,11,0,1,2,3,4,5,6,7,8,9},
    xtick=data,
    xticklabels={\hspace{-6pt}MB1,\hspace{1pt}MB2,0,1,2,3,4,5,6,7,8,9},
    x tick label style={font=\tiny, yshift=1pt},
    y tick label style={font=\tiny},
    ymin=0,
    ymax=2.0,
    grid=major,
    mark size=1.4pt,
    line width=0.9pt,
]

\addplot+[mark=*, solid] coordinates {
(10,0.122) (11,0.244) (0,0.379) (1,0.523)
(2,0.585) (3,0.652) (4,0.703) (5,0.780)
(6,0.847) (7,0.928) (8,1.043) (9,1.175)
};

\addplot+[mark=square*, solid] coordinates {
(10,0.223) (11,0.417) (0,0.663) (1,0.947)
(2,1.049) (3,1.163) (4,1.263) (5,1.375)
(6,1.474) (7,1.624) (8,1.783) (9,1.967)
};

\addplot+[mark=triangle*, solid] coordinates {
(10,0.120) (11,0.233) (0,0.420) (1,0.523)
(2,0.586) (3,0.649) (4,0.714) (5,0.770)
(6,0.844) (7,0.927) (8,1.043) (9,1.146)
};

\addplot+[mark=diamond*, solid] coordinates {
(10,0.111) (11,0.212) (0,0.328) (1,0.461)
(2,0.511) (3,0.568) (4,0.627) (5,0.680)
(6,0.735) (7,0.816) (8,0.918) (9,1.021)
};

\addplot+[mark=o, solid] coordinates {
(10,0.108) (11,0.207) (0,0.326) (1,0.453)
(2,0.512) (3,0.566) (4,0.619) (5,0.676)
(6,0.747) (7,0.811) (8,0.927) (9,1.054)
};

\addplot+[mark=+, solid] coordinates {
(10,0.114) (11,0.216) (0,0.337) (1,0.470)
(2,0.530) (3,0.584) (4,0.647) (5,0.711)
(6,0.777) (7,0.854) (8,0.946) (9,1.029)
};

\addplot+[mark=x, solid] coordinates {
(10,0.116) (11,0.216) (0,0.339) (1,0.474)
(2,0.540) (3,0.595) (4,0.656) (5,0.719)
(6,0.765) (7,0.837) (8,0.923) (9,1.027)
};

\addplot+[mark=pentagon*, solid] coordinates {
(10,0.136) (11,0.253) (0,0.413) (1,0.568)
(2,0.639) (3,0.700) (4,0.771) (5,0.837)
(6,0.896) (7,0.966) (8,1.069) (9,1.193)
};

\end{axis}

\end{tikzpicture}

\vspace{-0.01cm}

\begin{tikzpicture}
\begin{axis}[
    hide axis,
    width=\linewidth,
    height=0.25\linewidth,
    legend style={
        at={(0.5,0.5)},
        anchor=center,
        legend columns=4,
        font=\scriptsize,
        draw=black,
        fill=white,
        inner sep=3pt,
        column sep=8pt,
        row sep=2pt
    }
]

\addplot+[mark=*, solid] coordinates {(0,0)};
\addlegendentry{MAXN}

\addplot+[mark=square*, solid] coordinates {(0,0)};
\addlegendentry{10W}

\addplot+[mark=triangle*, solid] coordinates {(0,0)};
\addlegendentry{15W}

\addplot+[mark=diamond*, solid] coordinates {(0,0)};
\addlegendentry{30W\_ALL}

\addplot+[mark=o, solid] coordinates {(0,0)};
\addlegendentry{30W\_6C}

\addplot+[mark=+, solid] coordinates {(0,0)};
\addlegendentry{30W\_4C}

\addplot+[mark=x, solid] coordinates {(0,0)};
\addlegendentry{30W\_2C}

\addplot+[mark=pentagon*, solid] coordinates {(0,0)};
\addlegendentry{15W\_D}

\end{axis}
\end{tikzpicture}
\captionsetup{
    font=footnotesize,
    labelfont={bf,footnotesize}
}
\caption{Edge-side latency and energy per frame across TinyViT split points under different Jetson power and CPU configurations.}
\label{fig:latency_energy}
\vspace{-5mm}
\end{figure}

\begin{table}[t]
\centering
\renewcommand{\arraystretch}{1.25}
\setlength{\tabcolsep}{5pt}
\captionsetup{
    font=footnotesize,
    labelfont={bf,footnotesize}
}
\caption{Compressed representation size (MiB) at each TinyViT split point under different compression ratios.}
\label{tab:packet_size}

\begin{tabular}{lccc}
\hline
\rule{0pt}{2.6ex}
\textbf{Split Point} & \textbf{CR=0.05} & \textbf{CR=0.1} & \textbf{CR=0.25} \\
\hline
MB1         & 0.375 & 0.750 & 2.000 \\
MB2         & 0.188 & 0.375 & 1.000 \\
Block 0     & 0.188 & 0.375 & 1.000 \\
Blocks 1--6 & 0.062 & 0.125 & 0.312 \\
Blocks 7--9 & 0.125 & 0.250 & 0.625 \\
\hline
\end{tabular}
\end{table}

\vspace{-3mm}

\subsection{Quality-Constrained Edge Deployment Tradeoffs}
\label{sec:results_edge}
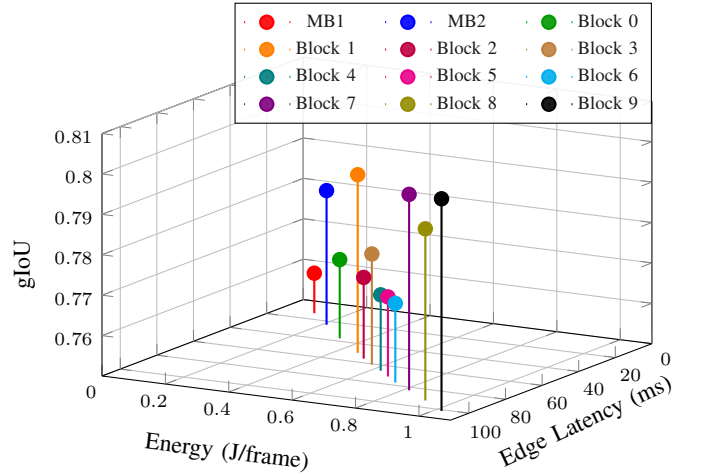
\begin{figure}[t]
\centering
\begin{tikzpicture}
\begin{axis}[
    width=\linewidth,
    height=0.72\linewidth,
    view={120}{20},
    xlabel={Edge Latency (ms)},
    ylabel={Energy (J/frame)},
    xlabel style={
        rotate=25,
        anchor=south,
        yshift=-8pt
    },
    ylabel style={
        rotate=-7,
        anchor=north,
        xshift=-8pt
    },
    zlabel={gIoU},
    xmin=0,
    xmax=110,
    ymin=0,
    ymax=1.10,
    zmin=0.75,
    zmax=0.81,
    xtick={0,20,40,60,80,100},
    ytick={0,0.2,0.4,0.6,0.8,1.0},
    ztick={0.76,0.77,0.78,0.79,0.80,0.81},
    grid=major,
    tick label style={font=\scriptsize},
    label style={font=\small},
    legend style={
        at={(1.0,1.15)},
        anchor=north east,
        legend columns=3,
        font=\scriptsize,
        fill=white,
        fill opacity=0.9,
        draw=black,
        /tikz/every even column/.append style={
            column sep=0.3cm
        }
    },
]

\addplot3[
    red,
    thick,
    ycomb,
    mark=*,
    mark size=2.5pt
] coordinates {
    (13.2,0.111,0.7599)
};
\addlegendentry{MB1}

\addplot3[
    blue,
    thick,
    ycomb,
    mark=*,
    mark size=2.5pt
] coordinates {
    (24.0,0.212,0.7832)
};
\addlegendentry{MB2}

\addplot3[
    green!60!black,
    thick,
    ycomb,
    mark=*,
    mark size=2.5pt
] coordinates {
    (36.9,0.328,0.7695)
};
\addlegendentry{Block 0}

\addplot3[
    orange,
    thick,
    ycomb,
    mark=*,
    mark size=2.5pt
] coordinates {
    (50.1,0.461,0.7941)
};
\addlegendentry{Block 1}

\addplot3[
    purple,
    thick,
    ycomb,
    mark=*,
    mark size=2.5pt
] coordinates {
    (55.5,0.511,0.7701)
};
\addlegendentry{Block 2}

\addplot3[
    brown,
    thick,
    ycomb,
    mark=*,
    mark size=2.5pt
] coordinates {
    (60.9,0.568,0.7774)
};
\addlegendentry{Block 3}

\addplot3[
    teal,
    thick,
    ycomb,
    mark=*,
    mark size=2.5pt
] coordinates {
    (66.3,0.627,0.7688)
};
\addlegendentry{Block 4}

\addplot3[
    magenta,
    thick,
    ycomb,
    mark=*,
    mark size=2.5pt
] coordinates {
    (71.5,0.680,0.7697)
};
\addlegendentry{Block 5}

\addplot3[
    cyan,
    thick,
    ycomb,
    mark=*,
    mark size=2.5pt
] coordinates {
    (77.0,0.735,0.7696)
};
\addlegendentry{Block 6}

\addplot3[
    violet,
    thick,
    ycomb,
    mark=*,
    mark size=2.5pt
] coordinates {
    (83.5,0.816,0.7985)
};
\addlegendentry{Block 7}

\addplot3[
    olive,
    thick,
    ycomb,
    mark=*,
    mark size=2.5pt
] coordinates {
    (92.3,0.918,0.7925)
};
\addlegendentry{Block 8}

\addplot3[
    black,
    thick,
    ycomb,
    mark=*,
    mark size=2.5pt
] coordinates {
    (101.3,1.021,0.8025)
};
\addlegendentry{Block 9}

\end{axis}
\end{tikzpicture}
\captionsetup{
    font=footnotesize,
    labelfont={bf,footnotesize}
}
\vspace{-6mm}
\caption{Quality-constrained edge design space across TinyViT split points at CR $=0.1$ under the 30\,W\_ALL configuration. Each point combines flood-adapted gIoU with edge-side latency and energy. A user-defined minimum gIoU requirement determines the quality-feasible subset, while the corresponding compressed representation sizes are reported in Table~\ref{tab:packet_size}.}
\label{fig:edge_tradeoff}
\vspace{-5mm}
\end{figure}

The comparatively compact flood-adapted accuracy range makes system cost an important factor in selecting among viable partitions. 
Fig.~\ref{fig:latency_energy} shows that deeper TinyViT partitions progressively increase edge-side latency and energy, while Table~\ref{tab:packet_size} shows that the communication footprint is non-monotonic because of the hierarchical feature dimensions. 
At CR $=0.1$, the compressed representation ranges from 0.750~MiB at MB1 to 0.125~MiB across Blocks~1--6, before increasing to 0.250~MiB across Blocks~7--9.

Fig.~\ref{fig:edge_tradeoff} combines flood-adapted gIoU with edge-side latency and energy under the 30\,W\_ALL configuration. 
Given a user-defined minimum quality requirement $Q_{\min}$, partitions satisfying $\mathrm{gIoU}\geq Q_{\min}$ form the quality-feasible set; the operating point can then be selected according to latency, energy, and communication costs.

For example, with $Q_{\min}=0.79$, Blocks~1, 7, 8, and~9 satisfy the quality requirement.
Block 1 provides the lowest latency, energy, and transmitted representation size among these quality-feasible points, achieving 0.7941 gIoU with 50.1~ms latency, 0.461~J per frame, and a 0.125~MiB compressed representation. 
Relative to the highest-quality Block~9 configuration, this corresponds to only a 0.84-point reduction in gIoU while reducing latency by 50.5\%, energy by 54.8\%, and transmitted feature size by 50\%. 
A stricter requirement of $Q_{\min}=0.80$, however, selects Block~9 as the only evaluated partition satisfying the quality constraint.

Platform configuration provides an additional deployment dimension. 
At Block~1, the 10\,W mode requires 142.8~ms and 0.947~J per frame, compared with 50.1~ms and 0.461~J under 30\,W\_ALL.
Thus, both the model partition and platform operating mode affect the resource cost of satisfying a given perception-quality requirement.

%% file: conclusion.tex
\vspace{-4mm}
\section{Conclusion and Future Work}

This work introduced \texttt{FloodReasonBench}, a task- and system-level benchmark for VLM reasoning segmentation in resource-constrained embodied flood response.
It combines \texttt{FloodResponseSeg}, a flood-specific reasoning-segmentation dataset, with systematic characterization of lightweight visual encoding, hierarchical split inference, compressed intermediate representations, and embedded execution. 
Our results show that flood-specific adaptation substantially improves task accuracy and that lightweight visual encoding preserves much of this performance. 
More importantly, while the generic pre-adaptation landscape exhibits pronounced partition-dependent accuracy variation, the flood-adapted target-workload landscape exhibits a substantially more compact accuracy range across partitions.
In this target deployment setting, multiple partitions therefore provide comparable task quality while incurring different latency, energy, and communication costs, making quality-constrained operating-point selection a central systems consideration.

Future work will extend \texttt{FloodResponseSeg} to additional response-relevant entities and scenarios, incorporate emerging reasoning-segmentation models and edge platforms, and evaluate end-to-end distributed and closed-loop embodied execution under varying network conditions.

